\documentclass[reprint,superscriptaddress,amsmath,amssymb,aps]{revtex4-2}
\usepackage{graphicx}
\usepackage{dcolumn}
\usepackage{bm}
\usepackage{hyperref}
\hypersetup{linktoc=page}
\hypersetup{
    colorlinks,
    citecolor=blue,    filecolor=blue,
    linkcolor=blue,     urlcolor=blue
}

\begin{document}
\title{Analysis of a kinetic model for electron heat transport in inertial confinement fusion plasmas}
\author{A. Chrisment}
\email{antoine.chrisment@u-bordeaux.fr}
\affiliation{CEA, DAM, DIF, F-91297 Arpajon Cedex, France}%
\affiliation{CELIA, Université de Bordeaux, CNRS, CEA, UMR 5107, F-33405 Talence, France}%
\author{P. Loiseau}%
\affiliation{CEA, DAM, DIF, F-91297 Arpajon Cedex, France}%
\affiliation{Université Paris-Saclay, CEA, LMCE, 91680 Bruyères-le-Châtel, France}
\author{J.-L. Feugeas}
\affiliation{CELIA, Université de Bordeaux, CNRS, CEA, UMR 5107, F-33405 Talence, France}%
\author{P.-E. Masson-Laborde}
\affiliation{CEA, DAM, DIF, F-91297 Arpajon Cedex, France}%
\affiliation{Université Paris-Saclay, CEA, LMCE, 91680 Bruyères-le-Châtel, France}
\author{J. Mathiaud}
\affiliation{CELIA, Université de Bordeaux, CNRS, CEA, UMR 5107, F-33405 Talence, France}%
\author{V. Tikhonchuk}
\affiliation{CELIA, Université de Bordeaux, CNRS, CEA, UMR 5107, F-33405 Talence, France}%
\affiliation{ELI-Beamlines Center, Institute of Physics, Czech Academy of Sciences, 25241 Dolni Brezany, Czech Republic}%
\author{Ph. Nicolaï}
\affiliation{CELIA, Université de Bordeaux, CNRS, CEA, UMR 5107, F-33405 Talence, France}%
\date{\today}

\begin{abstract}
To determine the electron heat flux density on macroscopic scales, the most widely used approach is to solve a diffusion equation through a multi-group technique. This method is however restricted to transport induced by temperature gradients without accounting for other sources of fast electrons because the electric field induced by the charge separation is indirectly treated. In addition, significant discrepancies are reported on the underlying distribution function when compared to complete kinetic calculations. These limitations motivate the research for alternative reduced kinetic models. The physical content of one of them is here deepened, its precision improved and the benefit of its usage compared to other models discussed.  
\end{abstract}
\maketitle

\section{\label{intro}Introduction}

One of the unresolved problems in inertial confinement fusion (ICF) research is an accurate description of energy transport on macroscopic scales. A difficulty comes from the necessity of treating the dynamics of fast electrons created by strong temperature gradients, parametric instabilities and laser resonant absorption in the presence of strong macroscopic inhomogeneities \cite{tikhonchuk2020progress, manheimer2021analytic}. The dedicated models must not only provide a closure to hydrodynamic equations - through mean fields, heat flux density and stress viscosity tensor \cite{braginskii1965transport} - but also be precise enough in order to account for kinetic processes at a macroscopic level.
The hydrodynamic approach reviewed by Brantov and Bychenkov \cite{brantov2013nonlocal, brantov2014nonlocal} achieves this program, but is limited to weakly inhomogeneous plasmas. It means that the variations of hydrodynamic profiles are supposed to be small compared to their mean values over space, without, however, any restriction on the steepness of their gradients with respect to the electron mean free path \cite{balescu1964kinetic}. This situation is encountered in the under-dense region where laser speckles produce localized small-amplitude perturbations of density and temperature in the plasma \cite{senecha1998temperature}. However, the approach of Refs. \cite{brantov2013nonlocal, brantov2014nonlocal} fails in the conduction zone or within the burning fuel. In these regions the existence of large variations of hydrodynamic profiles necessitates a kinetic treatment of electron transport. To accommodate the detail of the kinetic processes over macroscopic scales in a practical way, a 'reduced' kinetic model may be constructed. The utility of such an approach must strike a balance between numerical implementation efficiency and the requirements that both the electron distribution function and the self-consistent mean electric field be accurately represented. Such a model was introduced by Albritton \emph{et al.} \cite{albritton1986nonlocal} and has been recently studied by Del Sorbo \emph{et al.} \cite{del2015reduced} and Holec \emph{et al.} \cite{holec2018awbs} for the case of an unmagnetized plasma with a temperature gradient. These studies were limited, due to numerical issues, to temperature gradients that are not too steep. In this paper we report on an extension to steep temperature gradients thanks to a fast and robust scheme detailed in a companion paper \cite{Chrisment2022} and implemented in a code named JASMINE.

After having recalled the issues related to the problem of heat transfer in a zero-current-carrying plasma, the kinetic model is presented in Sec. \ref{sec2}. The hypotheses leading to the collision operator of Albritton \emph{et al.} \cite{albritton1986nonlocal} are highlighted and its accuracy is improved by a modification of the electron-electron collision frequency. For smooth temperature gradients analytic solutions are given in Sec. \ref{sec3}, and links are established between some of their features and the mathematical content of the model. Section \ref{sec4} is devoted to the case of a steep temperature gradient relevant to inertial confinement fusion. A conclusion is given in Sec. \ref{sec5}.   

\subsection{Coupling of the heat flux density to the electric field}\label{sec11}

We consider a fully ionized and unmagnetized plasma with a ionization number $Z$. The used reference frame - supposed Galilean - is the one moving at the mean velocity of ions, the inhomogeneities of which are supposed small compared to its mean value over space. In this reference frame, the electron mean velocity is $\mathbf{u}_e = \mathbf{0}$ as their macroscopic flow is attached to ions due to the ambipolar field. The electron temperature $T_e$ is monotonically decreasing from a hot region with temperature $T_H$ to a cold one with temperature $T_C$. The problem is treated in the quasi-static approximation as explained in Refs. \cite{luciani1985quasistatic, luciani1986resummation}. This means that we are examining the stationary kinetic state reached in the presence of fixed hydrodynamic constraints. The electron distribution function $f_e(\mathbf{r}, v, \boldsymbol{\Omega})$ thus obeys the stationary kinetic equation: 
\begin{align}
    v\,\boldsymbol{\Omega} \cdot \boldsymbol{\nabla} f_e & - e \mathbf{E}/m_e \cdot \left[ \boldsymbol{\Omega}\, \partial_v f_e + (\mathsf{I} - \boldsymbol{\Omega} \otimes \boldsymbol{\Omega})/v \cdot \partial_{\boldsymbol{\Omega}} f_e\right] \nonumber \\
    & = \mathcal{C}_{ei}[f_e,f_i] + \mathcal{C}_{ee}[f_e,f_e],
    \label{KineticEq}
\end{align}
where electron-ion, $\mathcal{C}_{ei}$,  and electron-electron, $\mathcal{C}_{ee}$, collision operators are Landau integrals \cite{landau1936kinetic}, $v$ is the absolute value of the electron velocity, $\boldsymbol{\Omega}$ is a unit vector in the velocity direction, $-e$ and $m_e$ are the electron charge and mass and $\mathbf{E}$ is the self-consistent electric field.

The problem of interest thus consists of finding the distribution function associated to the diffusion of electrons driven by a temperature gradient. The electron mean free path in the hot region is by a factor $(T_H/T_C)^2$ greater than in the cold one. This produces a forward electron current in the direction opposite to the temperature gradient, $-\boldsymbol{\nabla} T_e$. As electrons are charged, such a diffusion is responsible for a space charge creating an electric field $\mathbf{E}$. This electric field both slows down forward-moving electrons and drives a return current of electrons in the direction of $+\boldsymbol{\nabla} T_e$. 
As $\mathbf{u}_e = \mathbf{0}$ everywhere in the plasma, these fluxes of electrons are locally opposed and $\mathbf{E}$ takes a value such that the electric current density
\begin{align}
    \mathbf{j}_e = -en_e\mathbf{u}_e = -\frac{4\pi}{3} \int_0^{\infty} e v \mathbf{f}_1 v^2 dv,
    \label{Def j}
\end{align}
vanishes; $\mathbf{f}_1$ being the integral of $3/(4\pi) \boldsymbol{\Omega} f$ over the unit sphere $\mathbb{S}^2$. Because the mean free path $\lambda_{ei}$ is a monotonically increasing function of the velocity, $\mathbf{f}_1$ has only one zero - say, at $v=v_1$ - and the forward current in any point of the plasma is composed by the fastest electrons with velocities greater than $v_1$. To ensure $\mathbf{j}_e = 0$, the sum of $v^3 \mathbf{f}_1$ over velocities of the return current of electrons - from $0$ to $v_1$ - is thus the opposed to that of the forward current of electrons. Consequently, more electrons contribute to the return current. Being slower, the macroscopic gap to neutrality - over a distance larger than the Debye length - persists consistently with the existence of the mean electric field $\mathbf{E}$. However, this algebraic excess of electrons per unit of volume is negligible compared to the electron density $n_e$, which stays approximately equal to $Z n_i$ to high accuracy -  $n_i$ being the ion density. As the integrals of $v^3 \mathbf{f}_1$ over the velocities of forward and return electron currents compensate each other, it cannot be the case for  integrals of $v^5 \mathbf{f}_1$. This results in a nonzero heat flux density
\begin{align}
    \mathbf{q}_e = \frac{4\pi}{3}  \int_0^{\infty} \frac{m_e v^2}{2} v \mathbf{f}_1 v^2 dv,
    \label{Def q}
\end{align}
directed from the hot to the cold region. This phenomenological description emphasizes the importance of simultaneous evaluation of $\mathbf{q}_e$ and $\mathbf{E}$. 

The latter was not accounted for in the first formulation of the convolution integral approach introduced by Luciani {\it et al.} \cite{luciani1983nonlocal, mora1994nonlocal}, in which the heat flux density is expressed as:
\begin{align}
    \mathbf{q}_e(\mathbf{r}) = \int \mathsf{w} \big[\mathbf{r}, \mathbf{r}'; \lambda_d(\mathbf{r}') \big]\otimes \mathbf{q}_{\text{SH}}(\mathbf{r}') d\mathbf{r}', \label{eq LM}
\end{align}
where $\mathbf{q}_{\text{SH}}$ is the Spitzer-H{\"a}rm (SH) \cite{spitzer1953transport} heat flux density and $\lambda_d$ is the delocalization length. In Ref. \cite{luciani1983nonlocal} the latter is proportional to the average electron mean free path $\lambda_{ei}^T = 3 (\pi/2)^{1/2} \lambda_{ei}(v_T)$ \cite{braginskii1965transport}, where $\lambda_{ei}(v) = v/\nu_{ei}(v) = 4 \pi \epsilon_0^2 m_e^2 v^4/ (Z n_e e^4 \ln \Lambda)$ is the aforementioned velocity-dependent mean free path, $v_T=(k_B T_e/m_e)^{1/2}$ is the electron thermal velocity, $k_B$ is the Boltzmann constant, $\epsilon_0$ is the dielectric permittivity of vacuum and $\ln \Lambda$ is the Coulomb logarithm. 
In the limit of a small mean free path compared to the temperature scale length $L_T=T_e/\nabla T_e$, the SH formula must be recovered. Thus $w_{ij} \big[\mathbf{r}, \mathbf{r}'; \lambda_d(\mathbf{r}') \big]$ must tends toward the Dirac distribution $\delta_{ij} \delta (\mathbf{r} - \mathbf{r}')$ in the limit ${\rm Kn}=\lambda_{ei}^T/L_T \ll 1$. On the contrary, in the limit of a steep temperature gradient, ${\rm Kn} \gg 1$, the heat flux density cannot exceed the free-streaming value $n_e v_T k_B T_e$, directed opposite to the temperature gradient. These requirements are not too restrictive, and leave a large degree of freedom for choosing the shape of the kernel and an appropriate delocalization length. The exponential kernel introduced by Luciani {\it et al.} has been improved by Epperlein {\it et al.} \cite{Epperlein_1991} thanks to a Fourier analysis, and the electric field effect incorporated by Bendib {\it et al.} \cite{bendib1988improvement} through a modified delocalization length. Other kernels have been developed which are reviewed by Mora \emph{et al.} \cite{mora1994nonlocal} and recently, further analysed by Lu {\it et al} \cite{lu2021analytic}. 

For the purpose of determining the heat flux density, it is not necessary to use the expression \eqref{eq LM}. The kernel $\mathsf{w}$ can be considered as the Green function of a diffusion equation on $\mathbf{q}_e$, which may be directly solved. This equation is explicitly written in three dimensions at the end of Ref. \cite{mora1994nonlocal} but has never been implemented due to its intractability. To overcome this difficulty Schurtz, Nicolaï and Busquet (SNB) \cite{schurtz2000nonlocal} proposed to use the kernel $w_{ij} = 3/(4\pi) \Omega_i \Omega_j W(r)$, where $W$ is the kernel of Luciani \emph{et al.} in one dimension. By doing so they obtained a particularly convenient diffusion equation on $\mathbf{q}_e$ suitable for an efficient numerical implementation.
Instead of solving one equation for $\mathbf{q}_e$, they proposed a multigroup approach which consists of solving separately the transport problem for groups of electrons with different velocities. It thus relies on the kinetic detail of each term in the diffusion equation, and the method turns out to be equivalent to resolving a reduced system of kinetic equations on $\mathbf{f}_1$ and $f_0$, the isotropic part of the distribution function - defined as the integral of $f/(4 \pi)$ over $\mathbb{S}^2$. Despite the great progress it represents, the SNB model does not treat the electric field self-consistently but through the indirect way proposed by Bendib \emph{et al.} \cite{bendib1988improvement}. The convolution length is considered as the harmonic mean of a velocity-dependent mean free path and the electric stopping length $k_B T_e/|e\mathbf{E}|$, with the SH electric field \cite{spitzer1953transport}. Recent studies in one dimension by Sherlock \emph{et al.} \cite{sherlock2017comparison} and Brodrick \emph{et al.} \cite{brodrick2017testing} indicate acceptable accuracy of the SNB model compared to full kinetic computations of the heat flux density, but unsatisfactory behaviours of the distribution function. Such observations motivate us to further revision of reduced kinetic models devoted to electron heat transport.

Del Sorbo {\it et al.} \cite{del2015reduced} proposed a model with a self-consistent evaluation of the electric field. It is based on the electron-electron collision operator introduced by Albritton {\it et al.} \cite{albritton1986nonlocal}, and accounts for higher order an-isotropic components of the distribution function \cite{Minerbo_1978, Dubroca_2010}.
The distribution function was used for studying stability of the plasma in the transport region with respect to excitation of small-scale electron plasma and ion acoustic waves. However, significant differences from the full scale kinetic analysis were reported by Rozmus {\it et al.} \cite{rozmus2016resonance}. This indicates the need of extensive comparison of this reduced model to complete kinetic computations. We provide such a comparison in Secs. \ref{sec3} and \ref{sec4}, and analyse the benefit of using the Albritton {\it et al.} collision integral instead of the Bhatnagar, Gross, Krook (BGK) \cite{bhatnagar1954model} operator studied by Manheimer {\it et al.} \cite{manheimer2008development} and used in the SNB model.

\subsection{Importance of electron-electron collisions}\label{sec12}

For smooth temperature gradients, the electron distribution function is weakly perturbed and its isotropic part $f_0$ is close to the Maxwellian distribution function \cite{landau1981statistical}:
\begin{align}
    f_M = \frac{n_e}{(2\pi)^{3/2} v_T^3} e^{-v^2/2v_T^2},
    \label{fM}
\end{align}
where the electrostatic potential is accounted for implicitly in $n_e$ according to the Boltzmann distribution. The isotropic part $f_0$ being specified, it remains to analyse the anisotropic part of the distribution function, which mainly depends on collisions with momentum exchanges. For this reason the problem of determining both the heat current density $\mathbf{q}_e$ and the electric field $\mathbf{E}$ can be solved exactly in the high-$Z$ limit, where electron-ion collisions dominate. Indeed, due to the large ratio of masses between ion and electron, energy exchange in these collisions can be neglected, and the electron-ion Landau integral reduces to the Lorentz operator which describes elastic angular deflections. The electron-ion Lorentz operator, 
\begin{align}
    \mathcal{C}_{ei}(f_e) = \frac{\nu_{ei}}{2} (-\mathsf{L}^2) f_e, \label{eq Cei}
\end{align}
is proportional to the angular part of the Laplacian in velocity space:
\begin{align}
    - \mathsf{L}^2 = (\delta_{ij} - \Omega_i \Omega_j) \frac{\partial}{\partial \Omega_j} (\delta_{ik} - \Omega_i \Omega_k) \frac{\partial}{\partial \Omega_k},
\end{align}
where Einstein's summation convention over repeated indices is assumed. This operator, the negative of the square of the angular momentum operator in quantum mechanics, is the reason for using spherical coordinates $(v, \boldsymbol{\Omega})$ in the velocity space and decomposing $f_e$ on Maxwell multipoles eigenfunctions \cite{arnold2004lectures},
\begin{align}
    f_e  & = f_{l, i_1 \ldots i_l} P_{l, i_1 \ldots i_l} \nonumber\\
    & = f_0 + f_{1,i} \Omega_i + \frac{3f_{2, ij}}{2} \left(\Omega_i \Omega_j - \frac{\delta_{ij}}{3}\right) + \ldots
    \label{fExpansion}
\end{align}
with $- \mathsf{L}^2 P_{\ell, i_1 \ldots i_\ell} = -\ell(\ell+1) P_{\ell, i_1 \ldots i_\ell}$. 

As $Z$ decreases, the electron-electron collisions, which are solely responsible for energy exchange given the reduced form of the electron-ion collision operator used above, play an increasingly important role. Nevertheless, energy exchanges are not decisive in the case of a smooth temperature gradient because the departure of $f_0$ from $f_M$ is negligible. Concerning momentum exchanges, the electron-electron contribution is $Z$ times smaller than electron-ion collisions. An exact computation of $\mathbf{E}$ and $\mathbf{q}_e$ by using the electron-electron Landau integral relies on numerical simulations \cite{braginskii1965transport, spitzer1953transport}. 
Spitzer and H{\"a}rm \cite{spitzer1953transport} proposed to implicitly account for electron-electron collisions by introducing $Z$-dependent coefficients $\xi$ and $\zeta$ in formulas obtained in the Lorentz approximation ($Z\gg1$): 

\begin{align}
     \mathbf{E}_{\text{SH}} = - \frac{m_e v_T^2}{e} \left( \frac{\boldsymbol{\nabla} n_e}{n_e}  + \xi(Z) \frac{\boldsymbol{\nabla} T_e}{T_e} \right)  \label{ESH}
\end{align}
and 
\begin{align}
    \mathbf{q}_{\text{SH}} = - \frac{128 k_B n_e v_T^2} {3\pi\zeta(Z)\nu_{ei}^T} \boldsymbol{\nabla} T_e,    \label{qSH}
\end{align}
where $\nu_{ei}^T = v_T/\lambda_{ei}^T$ is the average electron-ion collision frequency. Being only dependent on the {\it local} values and shape of electron density and temperature, the transport regime these formulas describe is said to be local. The coefficients $\xi$ and $\zeta$ are obtained by interpolation, and in the notation of Spitzer and H{\"a}rm, $\xi = 1 + 3/2 \gamma_T/\gamma_E$ and $\zeta = (5/2 \epsilon \delta_T)^{-1}$. Their first order Padé approximations are \cite{epperlein1994effect}:
\begin{align}
    \xi(Z) = 1 + \frac{3}{2}\frac{Z+0.477}{Z+2.15}, \qquad \zeta(Z) = \frac{Z+4.2}{Z+0.24}
\end{align}

These empirical corrections are insufficient for steep temperature gradients where energy exchange due to electron-electron collisions must be explicitly accounted for. The local thermodynamic equilibrium hypothesis is no longer valid and there is a significant departure of $f_0$ from $f_M$. This means that energy reorganization among electrons becomes a decisive mechanism of their kinetics.    

\section{Kinetic model}\label{sec2}

\subsection{Approximate electron-electron collision operator}\label{sec21}

From a computational point of view, a complete treatment of electron-electron collisions using the Landau integral is prohibitively expensive on the temporal and spatial scales of interest for ICF. The approach adopted here consists of simplifying the Landau operator by considering collisions that involve only the strongly out-of-equilibrium electrons responsible for nonlocal effects. It corresponds to Chandrasekhar's picture \cite{chandrasekhar2005principles} of \emph{test} electrons interacting with a background of \emph{field} electrons in equilibrium. This approximation leads to the following electron-electron collision operator:
\begin{align}
    & \mathcal{C}_{ee}[f_e, f_M] =  \frac{\nu_{ee}}{2} \mathcal{H} \left( \frac{v}{v_T\sqrt{2}} \right) (-\mathsf{L}^2)f_e \nonumber\\
    & + \nu_{ee} v \partial_v \left[ \left( \frac{v}{v_T} \right)^2 \mathcal{G} \left( \frac{v}{v_T\sqrt{2}} \right) \mathcal{K} \right],
    \label{LinearizedCee}
\end{align}
where $\nu_{ee}$ is the velocity-dependent electron-electron collision frequency. The functions $\mathcal{H}$ and $\mathcal{G}$, first tabulated by Chandrasekhar, are defined from the error function - noted erf:
\begin{align}
\mathcal{H}(w) &= \frac{1}{2w^2} \left[ w \ \text{erf}^\prime (w) + (2w^2-1) \,\text{erf}(w) \right],\\
\mathcal{G}(w) & = \frac{1}{2w^2} \left[ \text{erf}(w) - w \ \text{erf}^\prime (w) \right],
\end{align}
and the function $\mathcal{K}$ is defined as follows:
\begin{align}
  \mathcal{K}= f_e + (v_T^2/v) \partial_v f_e. \label{defK}
\end{align}
Two further simplifications will be made in expression \eqref{LinearizedCee}  concerning the functions $\mathcal{H}$, $\mathcal{G}$ and $\mathcal{K}$.

\subsubsection{High velocity limit}\label{sec211}
Strongly-out-of-equilibrium electrons may be thought of as those with the highest velocities because the electron mean free path is proportional to $v^4$. Thus, we shall consider the high velocity limit of functions $\mathcal{H}$ and $\mathcal{G}$ at the lowest order:
\begin{align}
    \mathcal{H}\left( \frac{v}{v_T\sqrt{2}} \right) \simeq 1, \qquad \mathcal{G}\left( \frac{v}{v_T\sqrt{2}} \right) \simeq \left(\frac{v_T}{v} \right)^2. 
    \label{HVLChandraFunctions}
\end{align}

\begin{figure}[htp]
 \includegraphics[width=\linewidth]{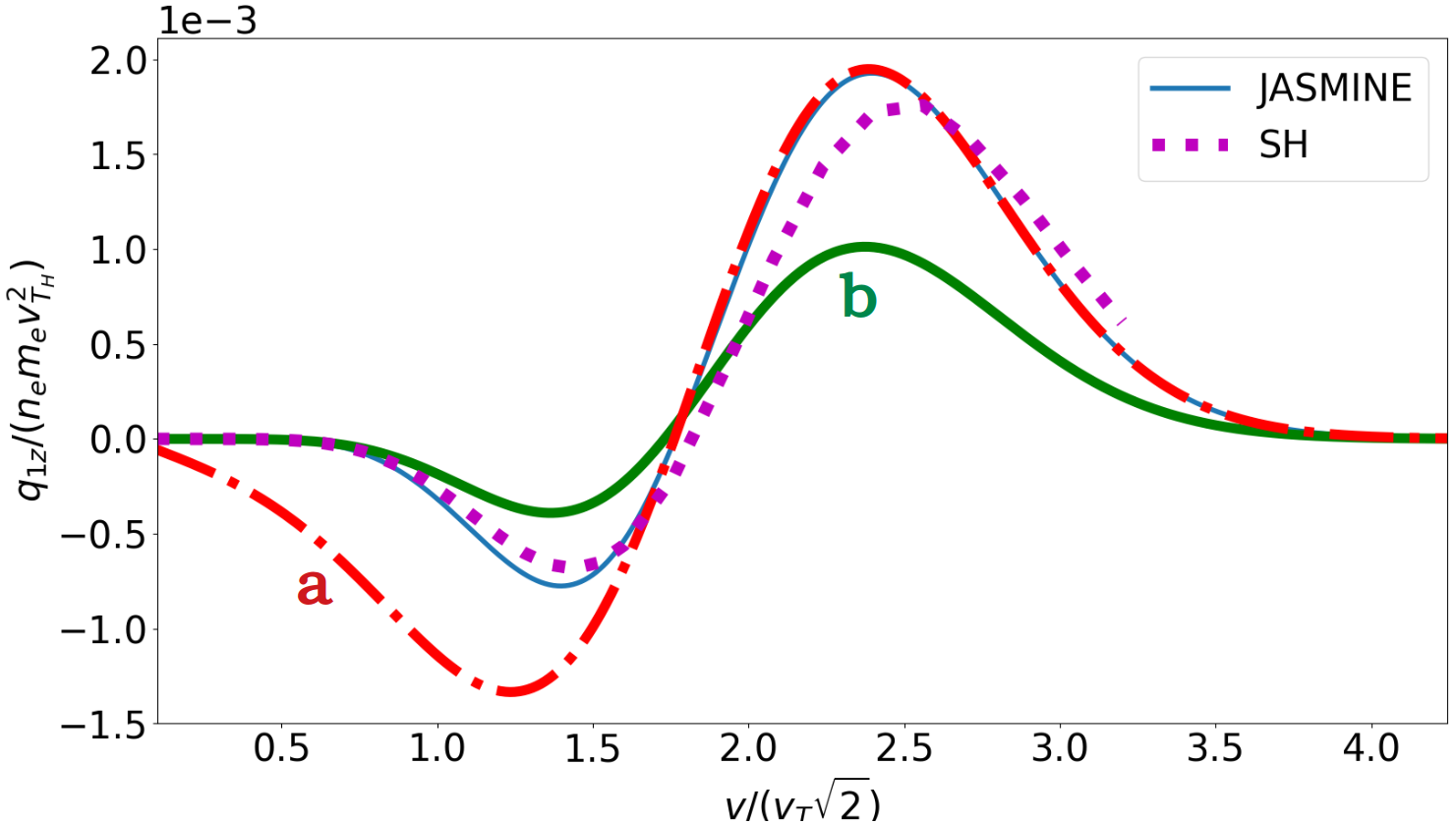}
 \caption{Effect of approximations \eqref{HVLChandraFunctions} and \eqref{eeCollRenormali} on the anisotropic part of electron distribution function $q_{1z} = (2\pi m_e/3) v^5f_{1z}$ in local regime for $Z=1$. The Spitzer-H{\"a}rm (SH) curve is the reference (purple, dotted); JASMINE (blue, solid) accounts for relations \eqref{HVLChandraFunctions}, \eqref{AWBSapprox} and \eqref{eeCollRenormali}. Curves (a, red, dot-dashed) and (b, green, solid) shows the results where approximations \eqref{HVLChandraFunctions} and \eqref{eeCollRenormali}, respectively, were omitted.}
 \label{Z1q1Local}
\end{figure}

\begin{figure}[htp]
 \includegraphics[width=\linewidth]{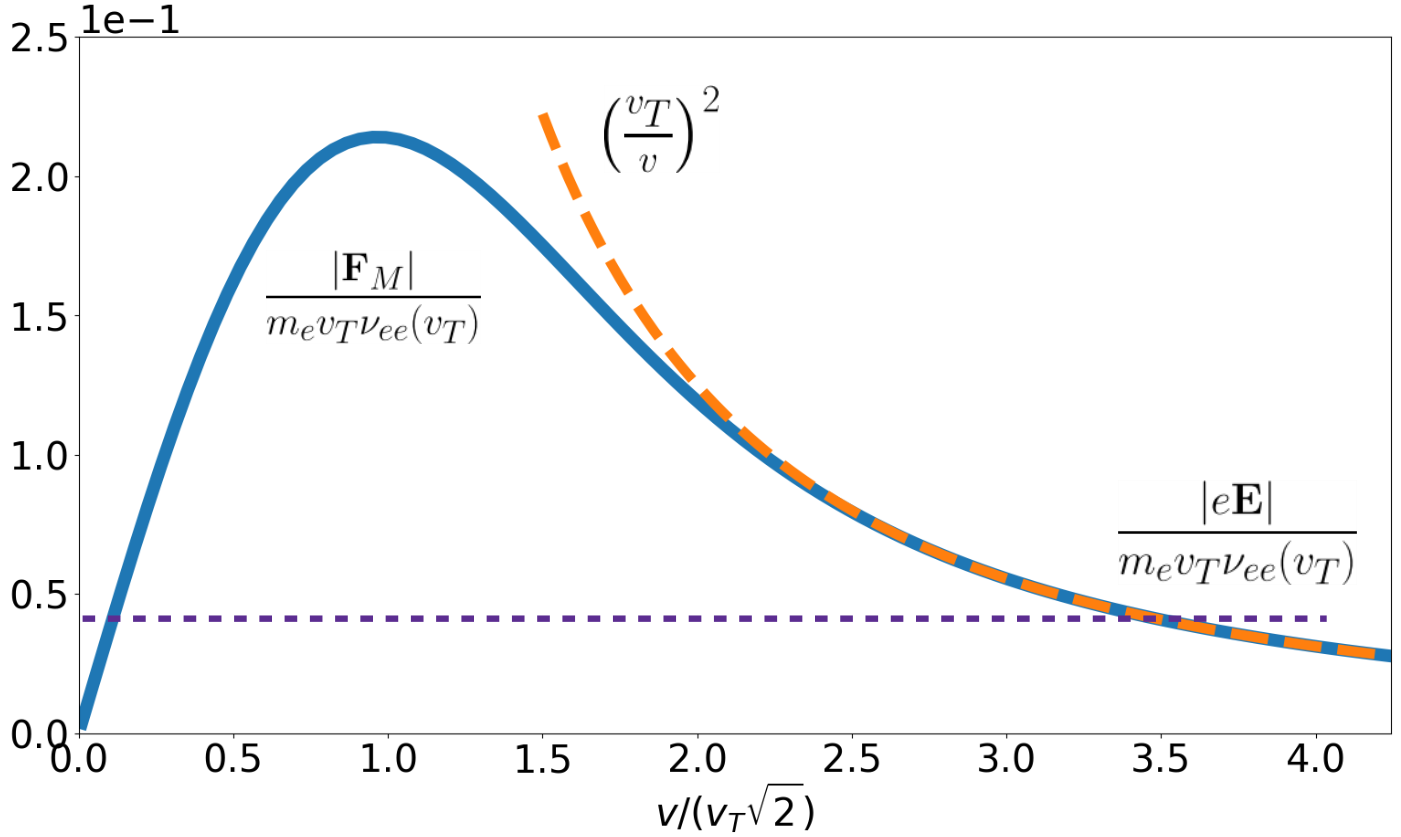}
 \caption{Graph of the normalized friction force (blue, solid), its high velocity approximation (orange, dashed) and the normalized electric force (violet, fine dashed). }
 \label{FrictionLorentzForces}
\end{figure}

The consequence of considering the high velocity limit of  $\mathcal{G}$ in \eqref{LinearizedCee} is the underestimation of the return electron current. This is shown in Fig. \ref{Z1q1Local}, where the normalized anisotropic part of electron distribution function $q_{1z} = (2\pi m_e/3)v^5 f_{1z}$ is plotted by either considering the complete Chandrasekhar (curve a, red, dot-dashed) functions or their high velocity expressions (curve JASMINE, blue, solid) \eqref{HVLChandraFunctions}. The reason for the difference can be found by analysing the norm of the dynamical friction force acting on a test electron with velocity $v$:
\begin{align}
    |\mathbf{F}_M| = m_ev_T \nu_{ee}(v_T) \mathcal{G} \left( \frac{v}{v_T\sqrt{2}}\right) \label{eqFM}
\end{align}
As shown in Fig. \ref{FrictionLorentzForces}, the high velocity approximation overestimates the friction force acting on test electrons by more than $10 \%$ for $v/(v_T \sqrt{2}) \lesssim 1.8$. This overestimation leads to a reduction of the return electron current. Such a reduction has been observed in all the cases we considered, in both local and non local regimes.  

On the contrary, considering the high velocity limit of $\mathcal{H}$ has no significant effect of the return current. In equation \eqref{LinearizedCee} this function appears as a factor in front of the Lorentz operator, and thus modulates the rate of the distribution function isotropization. Fortunately, as can be confirmed through numerical analysis, the anisotropic part of the distribution function is small at low velocities. Consequently, the error that arises due to taking the high velocity limit of the function $\mathcal{H}$ is negligible. 

By using expressions \eqref{HVLChandraFunctions} in the electron-electron collision operator \eqref{LinearizedCee} together with the Lorentz operator for electron-ion collisions \eqref{eq Cei}, we obtain the model used e.g by Gurevich and Istomin \cite{gurevich1979thermal} or by Luciani and Mora \cite{luciani1986resummation}. However, these papers do not address the problem of the numerical implementation efficiency of the collision operator, for which the double derivative in the radial part of \eqref{LinearizedCee} raises difficulties. For this purpose a simplification was proposed by Albritton \emph{et al.} \cite{albritton1986nonlocal}, which is examined in the next section.

\subsubsection{Equilibrium velocity diffusion}\label{sec212}

In expression \eqref{defK}, the second term is of second order in $v_T/v$, and should be neglected in accordance with expressions \eqref{HVLChandraFunctions} retained for Chandrasekhar functions. However, the function $\mathcal{K}$ will then lose the desirable property to be identically zero when $f_e=f_M$, for which the contribution of friction and diffusion in velocity space cancel each other. The simplification proposed by Albritton \emph{et al.} \cite{albritton1986nonlocal} consists of estimating the diffusion term $(v_T^2/v) \partial_v f_e$ by its equilibrium value $-f_M$, leading to: 
\begin{align}
    \mathcal{K} \simeq f_e - f_M.    \label{AWBSapprox}
\end{align}

Except the hottest region, each elementary volume of plasma contains an excess of suprathermal electrons. It means that for sufficiently high velocities $|\partial_v f_e| \leq |\partial_v f_M|$, and so expression \eqref{AWBSapprox} overestimates the diffusion term. The error can be estimated by considering \eqref{defK} as an equation on $f_e$. Since the latter is equal to $f_M$ for small velocities $v/v_T \ll 1$, an approximate solution can be written as \cite{krasheninnikov1993nonlocal}:
\begin{align}
    f_e = f_M + {\rm e}^{- v^2/2v_T^2} \int_0^{v} \mathcal{K}(w) \frac{w}{v_T^2} {\rm e}^{w^2/2v_T^2} dw.
\end{align}
Integrating by parts, we obtain:
\begin{align}
    \mathcal{K} =  f_e - f_M
     + {\rm e}^{- v^2/2v_T^2} \int_0^{v}  {\rm e}^{w^2/2v_T^2} [\partial_w \mathcal{K}]\, dw, 
     \label{KraSol}
\end{align}
where $\partial_w \mathcal{K} = \left( \mathcal{C}_{ee}[f_e, f_M] - \nu_{ee}/2 (- \mathsf{L}^2)f_e \right)/(w \nu_{ee})$ can be expressed from the kinetic equation \eqref{KineticEq}. As suggested by the factor $\exp(- v^2/2v_T^2)$, the integral term in \eqref{KraSol}, which is also equal to $v_T^2/v\, \partial_v (f_e-f_M)$, is small in the coldest region and in the high velocity domain. On the contrary, in the hottest region the departure of $f_e$ from $f_M$ due to the loss of the fastest electrons is small compared to the disequilibrium induced by their escape to colder regions. Consequently, expression \eqref{AWBSapprox} is violated for all velocities. This is numerically confirmed. In spite of this fact, we shall see that approximation \eqref{AWBSapprox} gives satisfying results. It means that simplification \eqref{AWBSapprox} partially compensates the omission of collisions between hot electrons, corresponding to a nonlinear term in the electron-electron collision integral.

\subsubsection{Corrected collision frequency}\label{sec213}

By taking into account simplifications \eqref{HVLChandraFunctions} and \eqref{AWBSapprox} we obtain the green curve (b) in Fig. \ref{Z1q1Local}. Both forward and return currents are approximately homothetic to those of the solution provided by Spitzer and H{\"a}rm in Tables I and II of Ref. \cite{spitzer1953transport}. Having fixed the contribution of the radial diffusion, their amplitudes can be only modified through the friction force \eqref{eqFM}, which is proportional to $\nu_{ee}$. Thus, we shall enhance forward and return current amplitudes by using a reduced electron-electron collision frequency $\nu_{ee}^*$:
\begin{align}
    \nu_{ee}^* = \phi(Z)\nu_{ee}.   \label{eeCollRenormali}
\end{align}
The correction factor $\phi(Z)$ is determined in Sec. \ref{sec31}. As a result of all approximations discussed above, the electron-electron collision operator \eqref{LinearizedCee} is:
\begin{align}
   \mathcal{C}_{ee}[f_e, f_M] \simeq \frac{\nu_{ee}^*}{2} (-\mathsf{L}^2) f_e +  \nu_{ee}^* v\partial_v(f_e-f_M). \label{eqCeestar}
\end{align}

\subsection{P1 closure}\label{sec22}

Energy exchange between electrons affects the isotropic part $f_0$ of the distribution function, whereas the anisotropic part is constrained by linear momentum exchanges. This simplified picture is completed by the coupling between the isotropic and anisotropic parts. As shown in \eqref{fExpansion}, the latter is \emph{a priori} composed by the sum of an arbitrarily large number of terms. However, the fact that the electric current \eqref{Def j} and heat flux \eqref{Def q} densities only depend on $\mathbf{f}_1$ suggests that the problem may be reduced to the study of a coupled system with unknown functions $f_0$ and $\mathbf{f}_1$. 

Such a system is obtained by projecting the kinetic equation \eqref{KineticEq} on $P_0 = 1$ and $\mathbf{P}_{1} = \mathbf{\Omega}$. These projections form a closed system on $f_0$ and $\mathbf{f}_1$ upon expressing the integral $\int_{\mathbb{S}^2} \mathbf{\Omega} \otimes \mathbf{\Omega} fd^2\Omega$ in terms of $f_0$ and $\mathbf{f}_1$ only. It can be achieved by imposing $f_{2,ij} = 0$:
\begin{align}
    \int_{\mathbb{S}^2} \left(\mathbf{\Omega} \otimes \mathbf{\Omega} - \mathsf{I}/3 \right) f d^2\Omega = 0
    \label{P1closure}
\end{align}
This procedure is referred as the P1 closure \cite{decoster1998modeling, Chrisment2022}. It yields the following system of coupled equations:
\begin{align}
    &\frac{v}{3} \boldsymbol{\nabla} \cdot \mathbf{f}_1 - \frac{e}{3m_ev^2} \mathbf{E} \cdot \partial_v v^2 \mathbf{f}_1 = \mathcal{C}_{ee}^0,    \label{eqf0} \\
   & v \boldsymbol{\nabla}f_0 - \frac{e}{m_e} \mathbf{E}\, \partial_v f_0 = \boldsymbol{\mathcal{C}}_{ee}^1 + \boldsymbol{\mathcal{C}}_{ei}^1
    \label{eqf1}
\end{align}
where the components of the electron-electron and electron-ion collision operators are given by:
\begin{align}
     &\mathcal{C}_{ee}^0 = v \nu_{ee}^* \partial_v(f_0-f_M),  \label{Collee0} \\
    & \boldsymbol{\mathcal{C}}_{ee}^1 = v \nu_{ee}^* \partial_v \mathbf{f}_1 - \nu_{ee}^* \mathbf{f}_1,  \label{Collee1} \\ 
    & \boldsymbol{\mathcal{C}}_{ei}^1 = - \nu_{ei}\mathbf{f}_1.
    \label{Collei1}
\end{align}

Relation \eqref{P1closure} sets the electron stress viscosity tensor to zero. Higher order terms $\ell \geq 3$ in expansion \eqref{fExpansion} are maintained to their initial values by the P1 closure. Those values are zeros in the diffusion problem we address, because the hydrodynamic constraints are entirely encoded in the initialisation $f_0 = f_M$.  In that context, the P1 closure is thus equivalent to seeking a solution of the form $f_0 + \mathbf{\Omega} \cdot \mathbf{f}_1$ \cite{Chrisment2022}. For a homogeneous plasma without mean fields the kinetic equation is reduced to $\partial_t f_e = \nu /2 (- \mathsf{L}^2)f_e$ with $\nu = \nu_{ei} + \nu_{ee}^*$, if energy exchanges are neglected. In that case all the modes $P_{\ell, i_1 ... i_\ell}$ relax as $[e^{-\nu t}]^{\ell(\ell+1)/2}$, and on a time of order $\nu^{-1}$, the modes $\ell \geq 2$ have already decreased at least ten times more than the mode $\ell=1$. Plasma inhomogeneities and energy exchange may prolong or reduce the lifespans of those modes but it is supposed, by making the P1 closure, that the aforementioned hierarchy between their relaxation remains valid. As it is shown below, the error created by omitting high order terms $\ell \geq 2$ is compensated by the renormalization of the electron-electron collision frequency \eqref{eeCollRenormali}.

The electric field entering in equations \eqref{eqf0} and \eqref{eqf1} is determined by the condition $\mathbf{j}_e = \mathbf{0}$. The Poisson equation cannot be used for this purpose because charge neutrality is imposed. Multiplying \eqref{eqf1} by $v^6$ and using the zero electric current condition, we obtain:
\begin{align}
    \mathbf{E} = -\frac{m_e}{e} & \left[ \int_0^{\infty} v^6 \boldsymbol{\mathcal{C}}_{ee}^1 dv - \int_0^{\infty} v^7 \boldsymbol{\nabla} f_0 dv  \right] \nonumber \\
    & \times \left[ \int_0^{\infty}   v^6 \partial_v f_0 dv  \right]^{-1}.
    \label{Efield}
\end{align}
Being proportional to $v^{-3} \mathbf{f}_1$, the contribution of $\boldsymbol{\mathcal{C}}_{ei}^1$ vanishes. The set of expressions \eqref{Def q}, \eqref{P1closure}, and \eqref{Efield} provides a complete kinetic closure to electron hydrodynamic equations. Equations \eqref{eqf0},  \eqref{eqf1} and \eqref{Efield} are numerically solved with the code JASMINE \cite{Chrisment2022}, and simulation results are discussed below.

The current densities \eqref{Def j} and \eqref{Def q} are driven by the competition between the advective and electric force terms acting on $f_0$. Indeed, let us write \eqref{eqf1} as  
\begin{align}
    \partial_v \mathbf{f}_1 + \frac{\alpha}{v} \mathbf{f}_1  = \frac{1}{ v \nu_{ee}^*} \left[ v \boldsymbol{\nabla} f_0 - \frac{e}{m_e} \mathbf{E} \partial_v f_0 \right]
\end{align}
where $\alpha(Z) = - [1 + Z/\phi(Z)]$. The left hand side is $v^{-\alpha} \partial_v v^\alpha \mathbf{f}_1$. Integrating between $v$ and $+\infty$ thus leads to:
\begin{align}
     \mathbf{f}_1 = -  \int_v^{\infty}  \left( \frac{w}{v}\right)^{\alpha} \left[ \frac{\boldsymbol{\nabla}f_0}{\nu_{ee}^*} - \frac{e\mathbf{E}}{m_e w \nu_{ee}^*}  \partial_w f_0 \right] dw
     \label{f1A}
\end{align}
It follows from this expression that spatial inhomogeneities define the value of the electric field. As shown in Fig. \ref{FrictionLorentzForces}, for steep temperature gradients the Lorentz electric force may overcome the friction force induced by collisions over a significant range of velocities. This is at the origin of numerical difficulties because the characteristics of the hyperbolic system of equations \eqref{eqf0}-\eqref{eqf1} do not have slopes with constant signs. Consequently, the integration on velocity cannot be performed from high to low values as done in Refs. \cite{del2015reduced, holec2018awbs}. This aspect is deepened and resolved in \cite{Chrisment2022}. 

Instead of using expression \eqref{f1A}, the heat flux density can be expressed by isolating $\mathbf{f}_1$ in Eq. \eqref{eqf1}:
\begin{align}
     &  \mathbf{q}_e = - \frac{2\pi m_e}{3\alpha} \int_0^{\infty} v^6 \partial_v  \mathbf{f}_1 dv \nonumber \\
    & - \frac{2(2\pi)^{1/2} Z m_e}{9 (\phi + Z) v_T^3 \nu_{ei}^T } \int_0^{\infty} v^8 \left[ v \boldsymbol{\nabla}f_0 - \frac{e}{m_e} \mathbf{E} \,\partial_v f_0 \right] dv
\end{align}
where we used $\nu_{ee}^* = 3 (\pi/2)^{1/2} \phi/Z (v_T/v)^3 \nu_{ei}^T$. Integrating by parts, the first term in the right hand side of this expression is equal to $6/\alpha \mathbf{q}_e$, such that
\begin{align}
    \mathbf{q}_e = - \frac{2 (2 \pi)^{1/2} Z m_e  }{9 (Z+7\phi) v_T^3 \nu_{ei}^T} \int_0^{\infty} v^8 \left[ v \boldsymbol{\nabla} f_0 -\frac{e}{m_e} \mathbf{E}\, \partial_v f_0\right] dv.
    \label{qA}
\end{align}
In the same way, we obtain for the electric current density:
\begin{align}
    \mathbf{j}_e = \frac{4(2\pi)^{1/2} Z e}{9(Z + 5\phi)v_T^3 \nu_{ei}^T} \int_0^{\infty} v^6 \left[ v \boldsymbol{\nabla} f_0 - \frac{e}{m_e} \mathbf{E}\, \partial_v f_0\right] dv.
    \label{elecCurrentDensity}
\end{align}
Since $\mathbf{j}_e=0$, the correction $\phi(Z)$ to the electron-electron collision frequency has no influence on the value of $\mathbf{E}$. This is a weak point of the P1 model. We examine it in the next section. 

\section{Local regime}\label{sec3}

\subsection{Condition to recover the Spitzer-H{\"a}rm electric field and heat flux density}\label{sec31}

Expression \eqref{Efield} for the electric field emphasizes the role of the first moment of the electron-electron operator. The local regime is characterized by a disequilibrium induced by the electron transport which remains small enough to ensure that $f_0$ is equal to $f_M$ to high accuracy. Substitution of $f_0$ by $f_M$ in \eqref{Efield} thus yields:
\begin{align}
    \mathbf{E} = - \frac{m_e v_T^2}{e} \left[ \frac{\boldsymbol{\nabla} n_e }{n_e} + \frac{5}{2}\frac{\boldsymbol{\nabla} T_e }{T_e} - \frac{(2\pi)^{3/2}}{48n_ev_T^5} \int_0^{\infty} v^6 \boldsymbol{\mathcal{C}}_{ee}^1 dv  \right].
\end{align}
The Spitzer-H{\"a}rm electric field \eqref{ESH} is thus recovered if, and only if
\begin{align}
    \int_0^{\infty} v^6 \boldsymbol{\mathcal{C}}_{ee}^1 dv = \frac{48n_e v_T^5}{(2\pi)^{3/2}} \left[ \frac{5}{2} - \xi(Z)  \right] \frac{\boldsymbol{\nabla} T_e}{T_e}.
\end{align}
However, in case of the operator given by Eq. \eqref{Collee1}, the left hand side of this relation is equal to zero. Indeed, the Lorentz part, $- \nu_{ee}^* \mathbf{f}_1$, gives no contribution because of the condition $\mathbf{j}_e = \mathbf{0}$, and, integrating the other contribution by parts, we find:
\begin{align}
    \int_0^{\infty} & v^6 v \nu_{ee}^* \partial_v \mathbf{f}_1 dv \nonumber \\
    & = - \frac{6 (2\pi)^{1/2} \phi(Z) v_T^3 \nu_{ei}^T}{Z} \int_0^{\infty} v^3 \mathbf{f}_1 dv = \mathbf{0}
\end{align}
Thus, the present model only allows recovery of the Lorentz electric field $\mathbf{E}_{\text{L}}$, corresponding to the expression \eqref{ESH} in the high-$Z$ limit, in which $\xi = 5/2$.
This observation has already been confirmed numerically by Holec \emph{et al.} \cite{holec2018awbs}.

\begin{figure}[htp]
 \includegraphics[width=\linewidth]{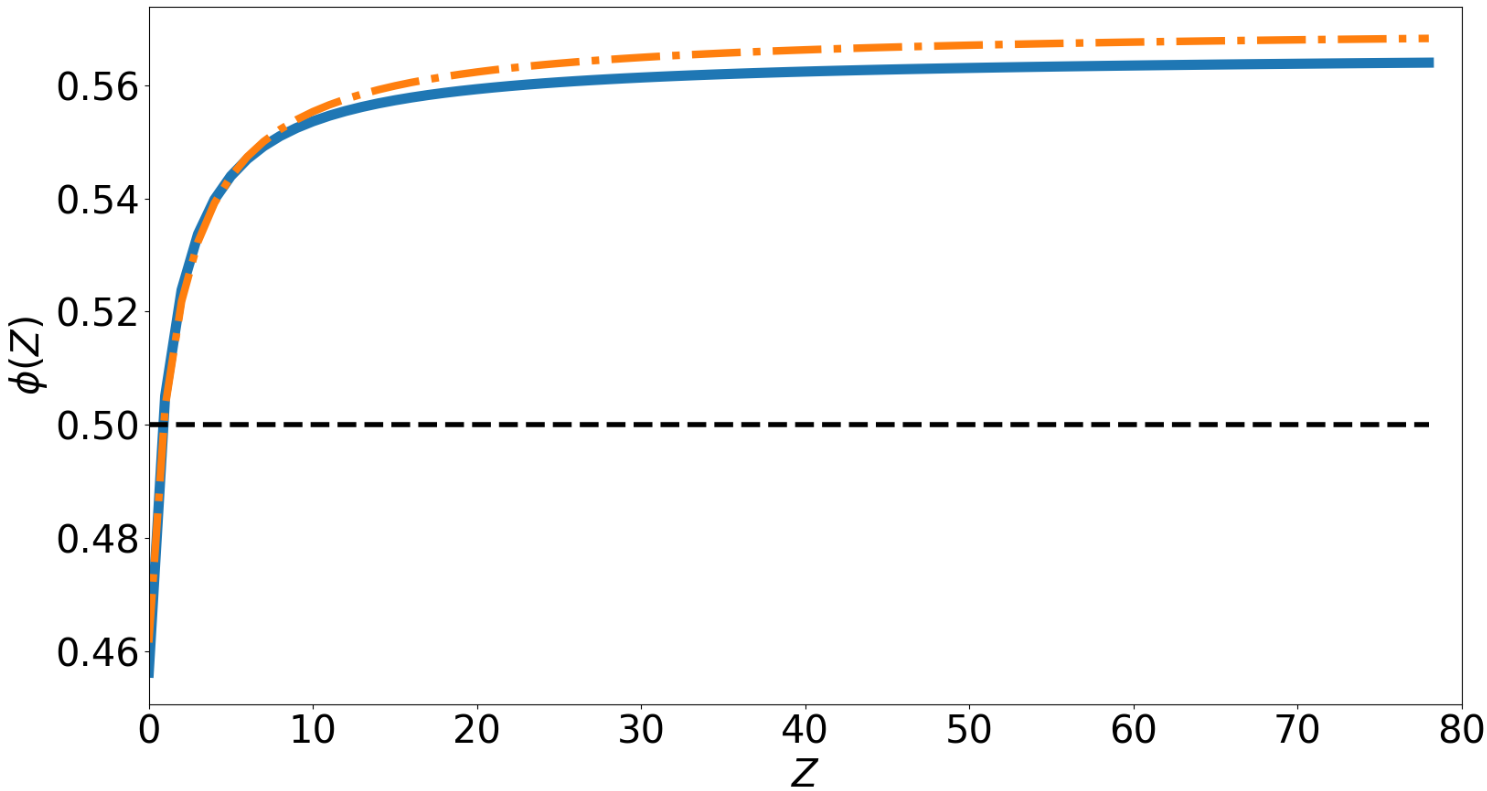}
 \caption{Correction factor $\phi$ for the electron-electron collision frequency given by the expression \eqref{phi} (blue, solid) and by the fit of Holec \emph{et al.} (orange, dashed-dot). The black dashed line is the value later authors choose to use in Ref. \cite{holec2018awbs} for numerical calculations. } \label{phiZ}
\end{figure}

Accounting for the Lorentz electric field, the heat flux density \eqref{qA} reads:
\begin{align}
    \mathbf{q}_e =  - \frac{128 k_B n_e v_T^2 Z}{3\pi (Z+7\phi) \nu_{ei}^T} \boldsymbol{\nabla} T_e
\end{align}
The above expression has to agree with the SH expression \eqref{qSH}, which implies that $7\phi/Z+1 = \zeta(Z)$. This condition provides the expression of the electron-electron correction factor:
\begin{align}
    \phi(Z) = \frac{Z}{7} [\zeta(Z) - 1] = \frac{3.96 Z}{7(Z+0.24)}.
    \label{phi}
\end{align}
As shown in Fig. \ref{phiZ}, this formula is consistent with the numerical fit proposed by Holec \emph{et al}: $1/2 + (0.59 Z - 1.11)/(8.37 Z + 5.15)$. The authors proposed to neglect the $Z$-dependent term of their fit in numerical calculations, but it corresponds to a systematic error of about $12 \%$.     

Without the correction \eqref{phi}, the electron heat flux density would be underestimated by 48\% for $Z=1$, by 5\% for $Z=50$ and by 4\% for $Z=79$.

\subsection{Effect of electron-electron collisions on $\mathbf{f}_1$}\label{sec32}

\begin{figure}[htp]
 \includegraphics[width=\linewidth]{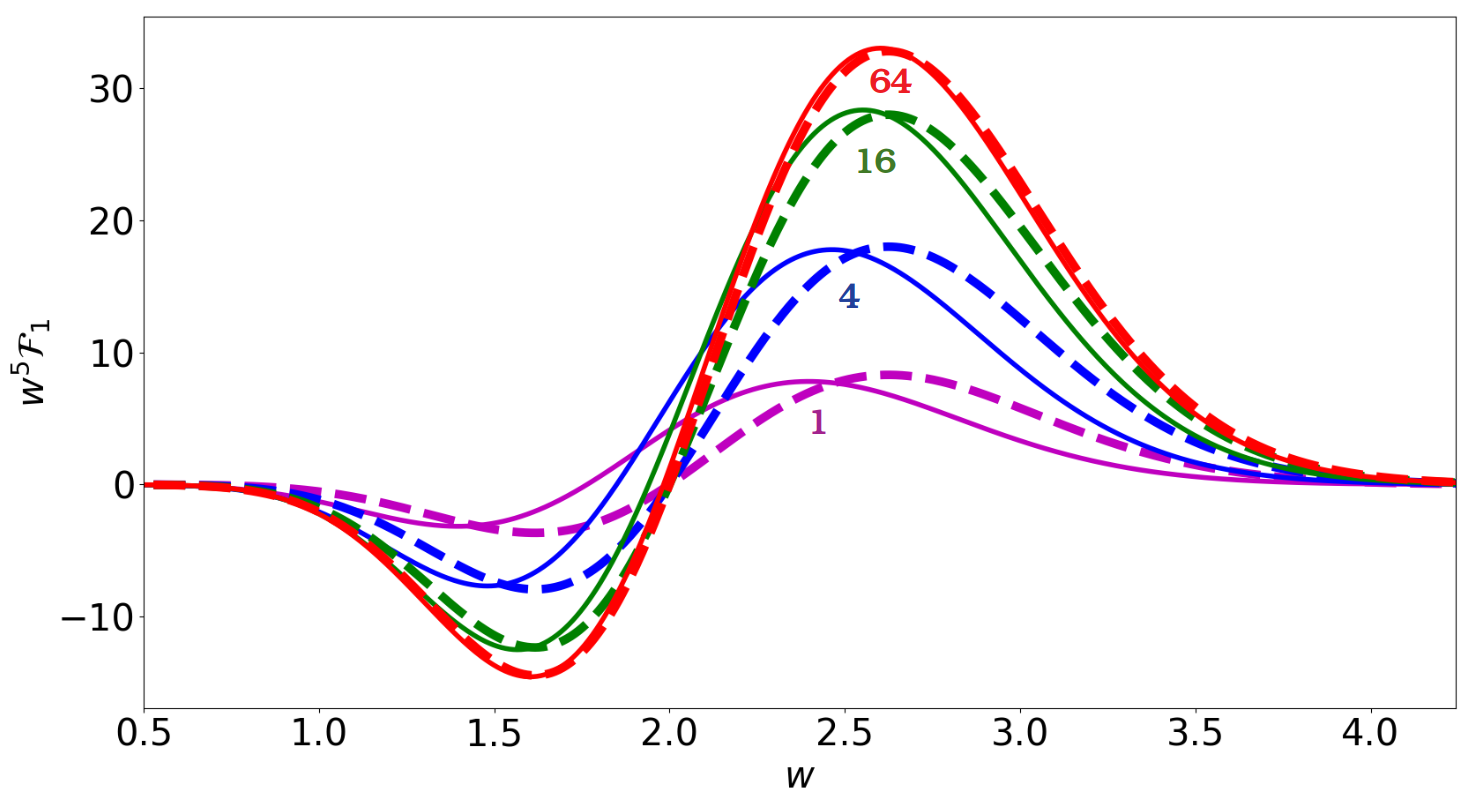}
 \caption{Graph of $w^5 \mathcal{F}_1$ with the function $\mathcal{F}_1$ given by expressions \eqref{F1A} (solid) and \eqref{f1L} (dashed), for different values of the ionization number: $Z=$ 1 (violet), 4 (blue), 16 (green) and 64 (red). }  
 \label{AtendstowardL}
\end{figure}

In the local regime, by substituting $f_0$ by the Maxwellian distribution function $f_M$ \eqref{fM}, the expression \eqref{f1A} can be written as:
\begin{align}
    & \mathbf{f}_{1,\text{A}} = - \frac{n_e}{3 \pi^2 v_T^2 \nu_{ei}^T} \mathcal{F}_{1,\text{A}} \left( \frac{v}{v_T \sqrt{2}}\right)\frac{\boldsymbol{\nabla} T_e}{T_e},
 \label{f1AL}
\end{align}   
where the function $\mathcal{F}_{1,\text{A}}$ is defined as:
\begin{align}
    & \mathcal{F}_{1,\text{A}} (w) = \frac{1 + \alpha}{w^\alpha}  \left[ 4 \Gamma \left(2 + \frac{\alpha}{2}; w^2 \right) - \Gamma \left( 3 + \frac{\alpha}{2}; w^2 \right)\right], 
    \label{F1A}
\end{align}
with $\Gamma$ the upper incomplete gamma function. It is worth mentioning that $\mathcal{F}_{1,\text{A}}$ is not defined at $w = 0$ because the first arguments $s = \{2,3\} + \alpha/2$ in $\Gamma$ functions are negatives for values of $Z$ greater than $\{ 1,2\}$. This results in a divergence of $\Gamma (s; w^2)$ as $- w^s/s$. In order to avoid the latter in the numerical implementation of $\mathcal{F}_{1,\text{A}}$, the lowest velocity of the mesh must be chosen not too low. As shown in Fig. \ref{AtendstowardL}, in the high $Z$ limit $\mathbf{f}_{1,\text{A}}$ tends toward the corrected Lorentz distribution function $\mathbf{f}_{1,\text{L}}$, defined by Eq. \eqref{f1AL} in substituting $\mathcal{F}_{1,\text{A}}$ for $\mathcal{F}_{1,\text{L}}$ defined as:
\begin{align}
       \mathcal{F}_{1,\text{L}}(w) = \frac{2w^4}{\zeta}(w^2 - 4)e^{-w^2} 
        \label{f1L}
\end{align}

\begin{figure}[htp]
 \includegraphics[width=\linewidth]{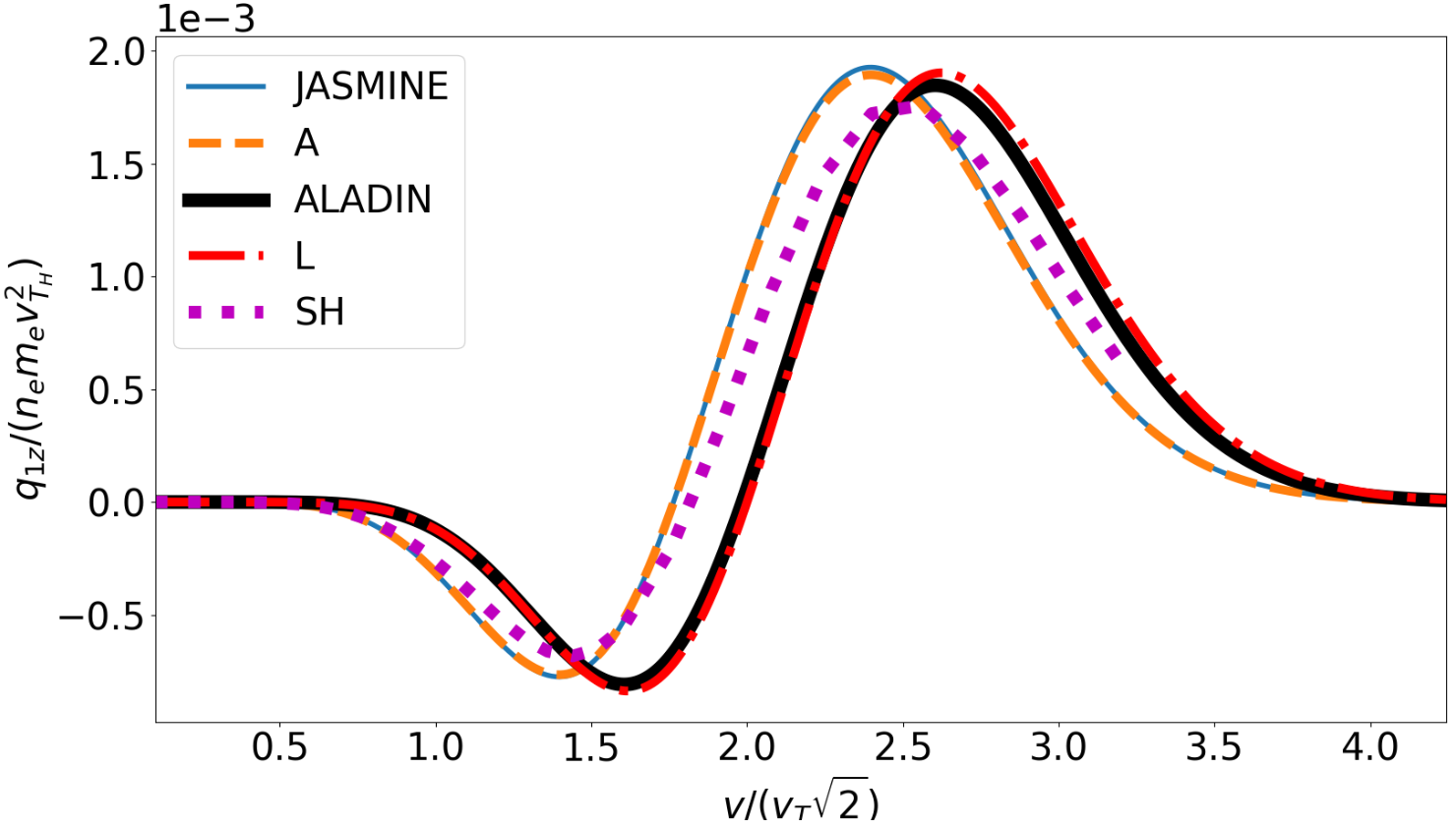}
 \caption{Effect of electron-electron collisions on the anisotropic part of electron distribution function $q_{1z} = (2\pi m_e/3) v^5\mathbf{f}_{1z}$ calculated with JASMINE code (blue, solid), ALADIN code (black, solid), SH tables (purple, dotted) and analytic solutions \eqref{f1AL} (orange, dashed) and \eqref{f1L} (red, dashed).  Local regime, $Z=1$.} \label{Z1q1Shift}
\end{figure}

The above expression \eqref{f1L} is found by solving the kinetic equation \eqref{eqf1} without $\boldsymbol{\mathcal{C}}_{ee}^1$ and by correcting the electron-ion collision frequency by a factor $\zeta(Z)$:
\begin{align}
    \boldsymbol{\mathcal{C}}_{ei}^1 + \boldsymbol{\mathcal{C}}_{ee}^1 = -\nu_{ei} \mathbf{f}_1 + \boldsymbol{\mathcal{C}}_{ee}^1 \simeq - \zeta(Z) \nu_{ei} \mathbf{f}_1
    \label{approxnuei}
\end{align}
Such an approximation for the collision term is used in many publications, e.g \cite{schurtz2000nonlocal, nikl2021implicit}, as it simplifies the numerical implementation. It is used in the code ALADIN \cite{aladin} where, however, a complete nonlinear expression is used for $\mathcal{C}_{ee}^0$ \cite{decoster1998modeling}. The simplification \eqref{approxnuei} has two consequences. The first one is the impossibility of recovering the SH electric field, as already pointed out in the previous section. The second one is a systematic error that appears in the position of $\mathbf{f}_1$ on the velocity axis for small values of $Z$. This fact is illustrated for $Z=1$ in Fig. \ref{Z1q1Shift}. It corresponds to the case of a plasma with constant density $n_e = 5 \times 10^{20}$ cm$^{-3}$ and  with temperature: 
\begin{align}
T_e(z) = T_H - \frac{(T_H - T_C)}{2} \tanh[s(z - z_0)],
\label{Tprofile}
\end{align}
where $T_H = 120$ eV, $T_C = 80$ eV, $z_0 = 425\,\mu$m and $s=9 \times 10^{-3}\,\mu$m$^{-1}$. The Knudsen number at the maximum of the gradient is Kn $= 7.3 \times 10^{-4}$. 

As shown in Fig. \ref{AtendstowardL}, it follows from Eq. \eqref{f1L} that $\mathbf{f}_{1,\text{L}}$ remains homothetic to itself for any variation of $Z$. This is not the case of $\mathbf{f}_{1,\text{A}}$, where the correction $\alpha(Z)$ does not enter as a simple multiplicative factor. This is responsible for a deformation corresponding to a shift toward low velocities compared to $\mathbf{f}_{1,\text{L}}$. For $Z=1$, this shift is slightly overestimated compared to the solution tabulated by Spitzer and H{\"a}rm \cite{spitzer1953transport}, $\mathbf{f}_{1, \text{SH}}$, defined by Eq. \eqref{f1A} in substituting $\mathcal{F}_{1, \text{A}}$ for $\mathcal{F}_{1, \text{SH}}$ defined as:
\begin{align}
    \mathcal{F}_{1, \text{SH}}(w) = - \left( 4 \frac{Z D_T \left( w \right)}{B} + 3 \frac{\gamma_T}{\gamma_E}\frac{Z D_E \left( w \right)}{A} \right)e^{-w^2},
    \label{f1SH}
\end{align}
where we used the notations introduced in Ref. \cite{spitzer1953transport}. The key feature to obtain the aforementioned effect is the differential nature of the operator describing energy exchange in $\boldsymbol{\mathcal{C}}_{ee}^1$. Using for the latter the algebraic BGK operator, $- \nu_{ee}^*(f_e-f_M)$, would not lead to any shift, but only to a modification of the $Z$-correction involved in the electron-electron collision frequency: $\nu_{ee}^* = \eta(Z) \nu_{ee}$ where $\eta(Z) = Z [\zeta(Z) - 1]/2 = 7 \phi(Z)/2$.
In the nonlocal regime this low $Z$ shift persists and has already been reported by Holec \emph{et al.} \cite{holec2018awbs}. 

\section{Nonlocal regime}\label{sec4}

Results obtained with the reduced model described in Sec. \ref{sec22} and implemented in our code JASMINE are compared to the numerical simulations performed with ALADIN, the full kinetic code OSHUN \cite{tzoufras2013multi, joglekar2018validation} and through the SNB model. 

OSHUN solves the non-stationary kinetic equation with electron-electron and electron-ion Landau integrals, by expanding the distribution function on spherical harmonics. The number of terms in the expansion is increased until reaching convergence. The SNB model consists of a first order perturbation resolution \cite{schurtz2000nonlocal} of the system \eqref{eqf0}-\eqref{eqf1} with the BGK operator for electron-electron energy exchange:  $\mathcal{C}_{ee}^0 = - \eta(Z) \nu_{ee} (f_0 - f_M)$ and $\boldsymbol{\mathcal{C}}_{ei}^1 + \boldsymbol{\mathcal{C}}_{ee}^1 =  - \zeta(Z) \nu_{ei} \mathbf{f}_1$ \eqref{approxnuei}. Zero-order solutions are given by expressions in the local regime: $f_0 = f_M$ \eqref{fM}, $\mathbf{f}_1 = \mathbf{f}_{1, \text{L}}$ \eqref{f1L} and $\mathbf{E} = \mathbf{E}_{\text{L}}$. The perturbation $\delta f_0 = f_0 - f_M$ obeys the following equation:
\begin{align}
    - \boldsymbol{\nabla} \cdot \frac{v^2}{3 \nu_{ei}^E }\boldsymbol{\nabla} \delta f_0 + \eta(Z) \nu_{ee} \delta f_0 = -v \boldsymbol{\nabla} \cdot \frac{v f_M}{3 \zeta(Z) \nu_{ei}}  \frac{\boldsymbol{\nabla} T_e}{T_e}, \label{eqDeltaf0} 
\end{align}
and $\delta \mathbf{f}_1 = \mathbf{f}_1 - \mathbf{f}_{1,\text{L}}$ is deduced from the relation $\delta \mathbf{f}_1 = - v/\nu_{ei}^E \boldsymbol{\nabla} \delta f_0$. The effective electron-ion collision frequency,  $\nu_{ei}^E = \zeta(Z) \nu_{ei} + 2 e |\mathbf{E}_\text{SH}|/(m_e v)$, indirectly accounts for the electric field effect on these perturbations. 

The considered case is representative of the conditions appearing in the conduction region during a laser-driven implosion process.  Initially the electron temperature is given by the expression \eqref{Tprofile} with $T_H = 1.8$ keV, $T_C = 0.2T_H$, $z_0 = 1200\,\mu$m and $s=2.6 \times 10^{-2}\,\mu$m$^{-1}$. The density profile is chosen such as the electron pressure $n_e k_B T_e$ is constant over the entire computational box and $n_e$ is equal to $n_H=1\times 10^{21}$ cm$^{-3}$ in the hot region. In kinetic simulations the initial heat flux density is zero. After a transient growth it is established in several collision times - a few picoseconds - and then follows, together with the temperature, a quasi-static evolution at the collision timescale according to $(3/2)\partial_t  n_e k_B T_e + \partial_z q_{ez} = 0$. At a given time of this evolution, the temperature profile is extracted from OSHUN and used in JASMINE and SNB calculations. 
The density, which evolves on the nanosecond timescale, can be considered unchanged. The effective ionization number is $Z=1$ and for the temperature profile at the considered time the Knudsen number at the maximum of the gradient is Kn = $\lambda_{ei}^T/L_T = 2.4 \times 10^{-2}$.

\subsection{Comparison of heat flux densities and electric fields}\label{sec41}

\begin{figure*}
 \includegraphics[width=12cm]{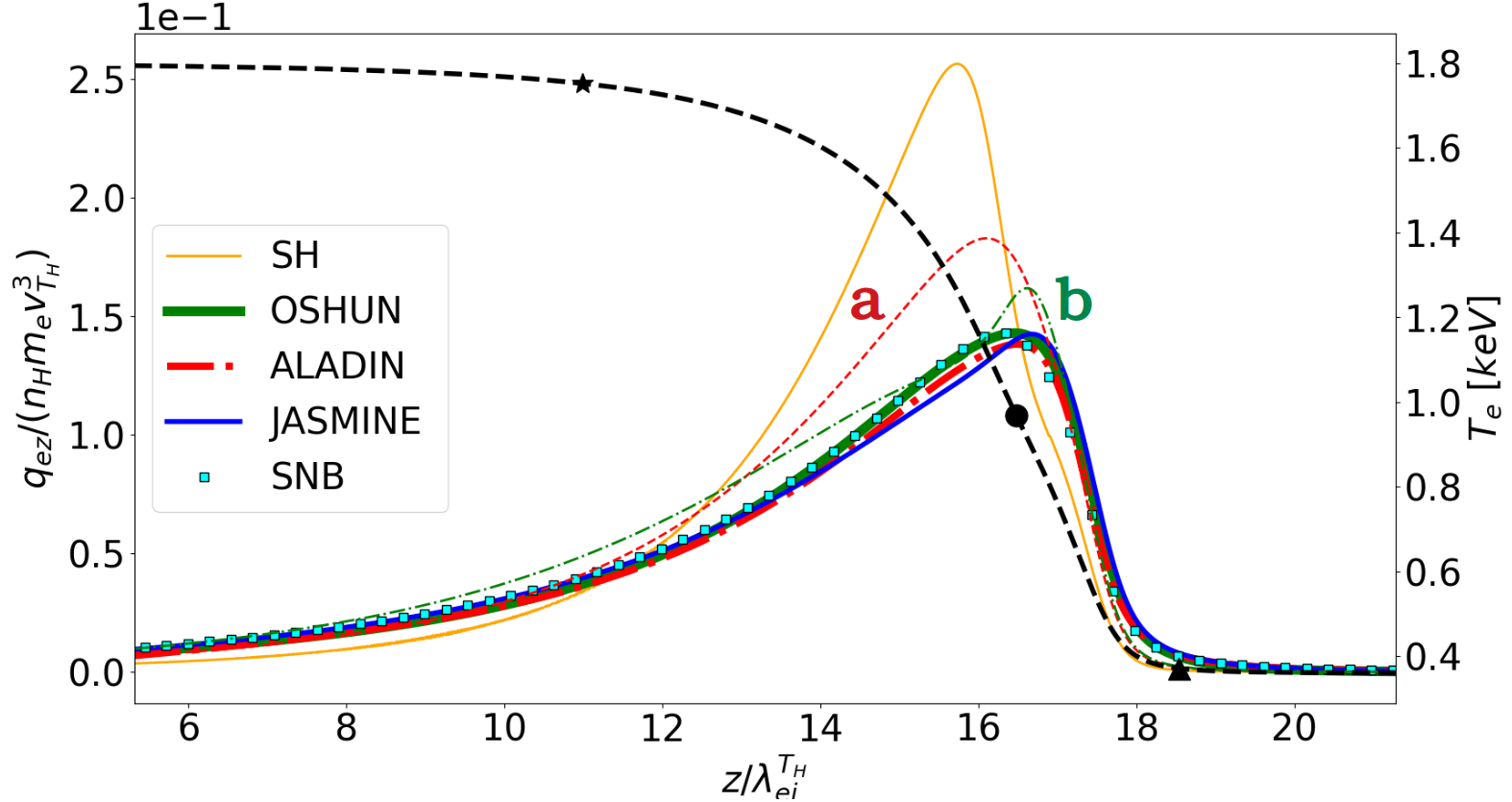}
     \caption{Heat flux $q_{ez}$ (left axis) and temperature $T_e$ (right axis) distribution in the simulations with JASMINE (blue, solid), ALADIN (red, dashed-dot), OSHUN (green, solid) and SNB (cyan, square). The orange solid line shows the tabulated Spitzer-H{\"a}rm (SH) heat flux density, fine color lines show JASMINE calculations (a) with the SH electric field in red dashed and (b) with approximation \eqref{approxnuei} in green dashed-dot. The heat flux density is normalized by the free streaming flux $n_H m_e v_{T_H}^3$ in the hot region. Star, circle and triangle on the temperature profile mark the positions where distribution functions are shown. The plasma size is $2000\,\mu$m, $T_H = 1.8$ keV, $T_C = 0.2T_H$, $Z =1$, and $\lambda_{ei}^{T_H}=73\,\mu$m. The time of comparison is 22 ps.
     }  \label{qNonlocal}
\end{figure*}

Heat flux densities are compared in Fig. \ref{qNonlocal}. JASMINE and SNB calculations agree with the kinetic results within a maximum difference of 10\%. In the steepest region, kinetic estimations of JASMINE, SNB, ALADIN and OSHUN are approximately 50\% smaller than the SH one \eqref{qSH}. This inhibition, together with the preheat at the foot of the temperature gradient due to electrons coming from the hot region, are signature of the nonlocal regime. The choice of an approximation for the electron-electron collision operator is important. If the term $\boldsymbol{\mathcal{C}}_{ee}^1$ is neglected, a more significant error is obtained in JASMINE calculation (green dashed line b) in Fig. \ref{qNonlocal}). However, at the time of comparison, the temperature profiles in OSHUN and ALADIN are slightly different. By using the latter, the results provided by JASMINE are about $5 \%$ larger, at the maximum of the heat flux density, than those presented in Fig. \ref{qNonlocal}. This quantifies the precision below which errors between codes cannot be compared.    

\begin{figure*}
     \includegraphics[width=12cm]{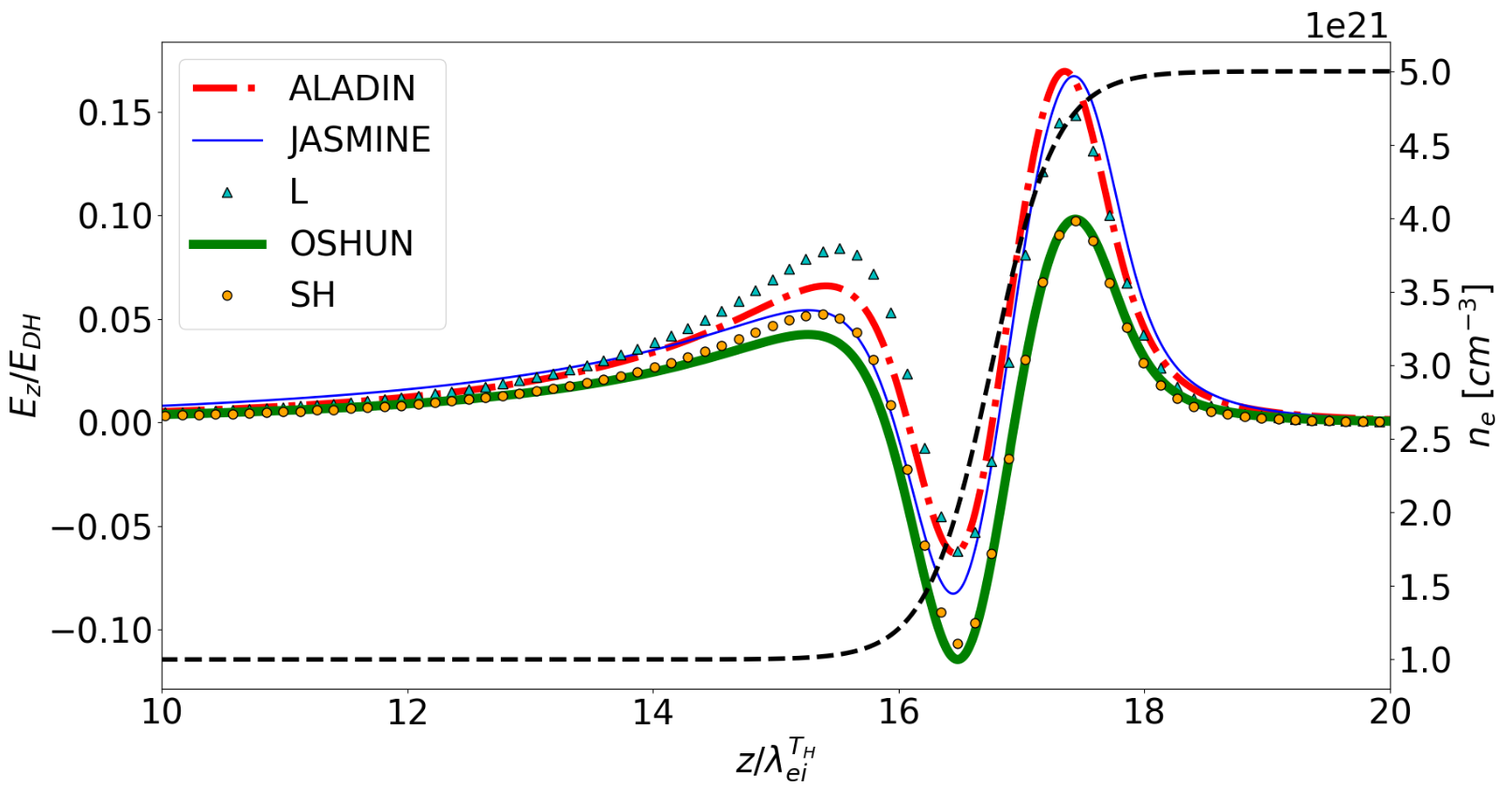}
     \caption{Electric field $E_z$ (left axis) and density $n_e$ (right axis). Cyan triangles show the analytic Lorentz (L) electric field. The electric field is normalized by the Dreicer field $E_{DH} = m_e v_{T} \nu_{ee}(v_T)/e$ in the hot region. In the SNB model the electric field is not self-consistently calculated. Other parameters and notations are the same as in Fig. \ref{qNonlocal}.} \label{ENonLocal}
\end{figure*}

Electric fields are compared in Fig. \ref{ENonLocal}. The OSHUN result is very close to the SH electric field \eqref{ESH}. ALADIN and JASMINE electric fields are larger and close to the Lorentz one. This means that electric field is much less affected by the steepness of the temperature gradient and remains close to its value predicted by the local regime expression. It explains the success of the SNB model \cite{schurtz2000nonlocal} because the SH expression \eqref{ESH} of the electric field is used in $\nu_{ei}^E$ for estimating 
\begin{align*}
    \mathbf{q} - \mathbf{q}_{\text{SH}} = \delta \mathbf{q} = - \frac{2 \pi m_e}{3} \int_0^{\infty} \frac{v^6}{\nu_{ei}^E} \boldsymbol{\nabla} \delta f_0 dv,
\end{align*}
instead of the Lorentz expression corresponding to the zero-order solution of the kinetic system considered by the SNB model.

Keeping this observation in mind, we investigate the consequences of using the SH electric field \eqref{ESH} while solving the system of equations \eqref{eqf0} and \eqref{eqf1}. As it does not ensure the condition $\mathbf{j}_e = \mathbf{0}$, it is responsible for a nonzero electron flux in the direction of $ \boldsymbol{\nabla} T_e$ everywhere in the plasma. However, the corresponding velocity $\mathbf{u}_e=-\mathbf{j}_e/e n_e$ related to this flux is negligible compared to $v_T$. Thus, one may neglect the shift it induces in formula \eqref{fM} defining the local equilibrium distribution function. An order of magnitude estimate can be obtained from \eqref{elecCurrentDensity} in the local regime:
\begin{align}
    \mathbf{j}_e = \frac{32 e^2 n_e Z}{3\pi m_e ( Z + 5 \phi) \nu_{ei}^T} \left[ \mathbf{E} + \frac{m_ev_T^2}{e} \left(  \frac{\boldsymbol{\nabla} n_e}{n_e} +\frac{5}{2} \frac{\boldsymbol{\nabla} T_e}{T_e} \right)\right] \nonumber
\end{align}
Substituting $\mathbf{E}_{\text{SH}}$ \eqref{ESH} in this formula leads to:
\begin{align}
   \frac{|\mathbf{u}_e|}{v_T} = \frac{120 [5/2 - \xi(Z)]}{1 + 5\phi(Z)/Z} \text{Kn} \nonumber
\end{align}
For any $Z$ the ratio $|\mathbf{u}_e|/v_T$ is three times smaller than $\text{Kn}$, and so is negligible compared to $1$. This statement holds if, in place of $\phi$ \eqref{phi}, the new correction $(Z/7)\,[\zeta(7/2 - \xi) - 1] $, required to recover the SH heat flux density in the local regime is used.
By using $\textbf{E}_{\text{SH}}$ together with this new correction JASMINE provides the heat flux density plotted with the red dashed curve (a) in Fig. \ref{qNonlocal}.  

It may be observed that the electric field changes sign. The two terms in \eqref{ESH} describe the space charge induced by the gradients of density and temperature. For a constant pressure, which implies $\boldsymbol{\nabla} n_e / n_e = -\boldsymbol{\nabla} T_e / T_e$, the electric field can be either expressed through the gradient of density or temperature only and thus keeps the same sign. As density and temperature inhomogeneities relax on different timescales, the pressure is not constant in the case shown in Fig. \ref{ENonLocal}, and its gradient is responsible for the negative part of the electric field. It accelerates electrons in the direction opposite to the temperature gradient, thus enhancing the deformation of the distribution function in the cold region.

\subsection{Comparison of distribution functions}\label{sec42}

\begin{figure*}
     \includegraphics[width=12cm]{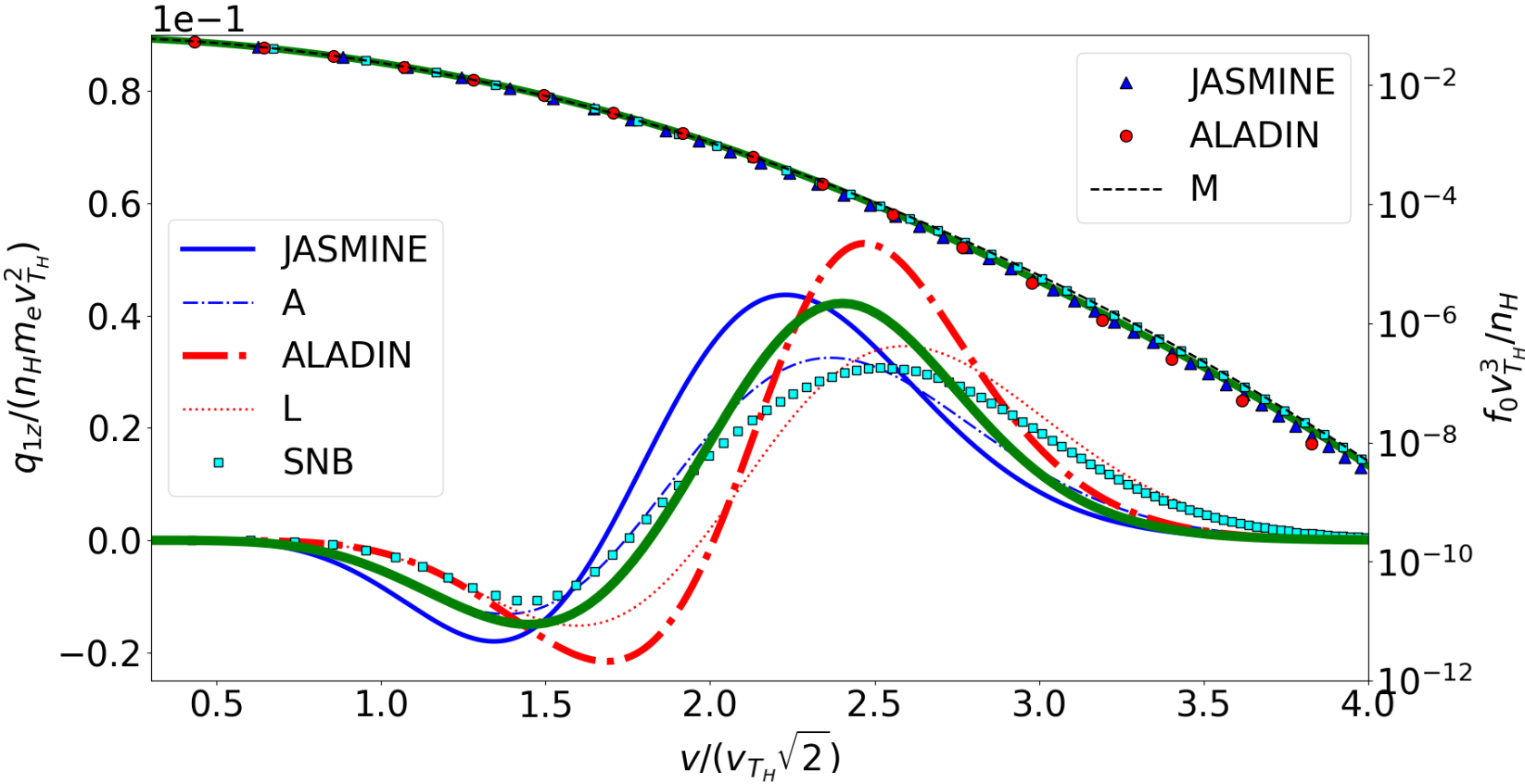}
     \caption{Comparison of the distribution functions $f_0$ (right axis) and $q_{1z} = (2\pi m_e/3) v^5f_{1z}$ (left axis) calculated with JASMINE, ALADIN, OSHUN and SNB codes at the point of hot region marked by a star in Fig. \ref{qNonlocal} ($z/\lambda_{ei}^{T_H} \simeq 11$). Black dashed line (right axis) shows the Maxwellian (M) distribution function. Fine lines show the heat flux densities integrand $q_{1z}$ associated to the analytic solution $f_{1,\text{A}}$ \eqref{f1A} in dashed-dot blue and $f_{1,\text{L}}$ \eqref{f1L} in dotted red. Other parameters and notations are the same as in Fig. \ref{qNonlocal}.}  \label{q1NONLOCAL1}
\end{figure*}

\begin{figure*}
     \includegraphics[width=12cm]{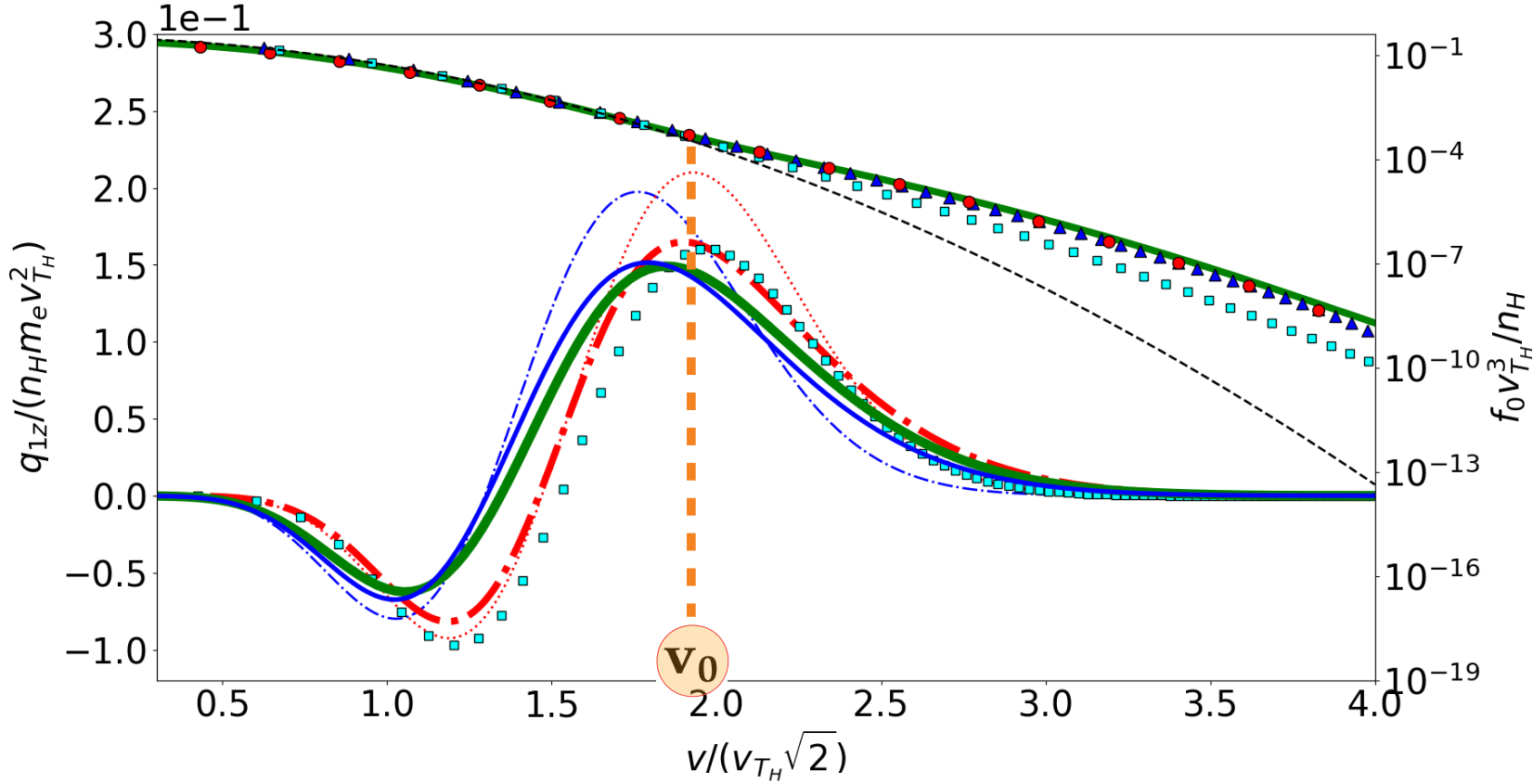}
     \caption{Same as in Fig. \ref{q1NONLOCAL1} for the point at the middle of the temperature gradient marked by a circle in Fig. \ref{qNonlocal} ($z/\lambda_{ei}^{T_H} \simeq 16.5$).} \label{q1NONLOCAL2}
\end{figure*}

\begin{figure*}
     \includegraphics[width=12cm]{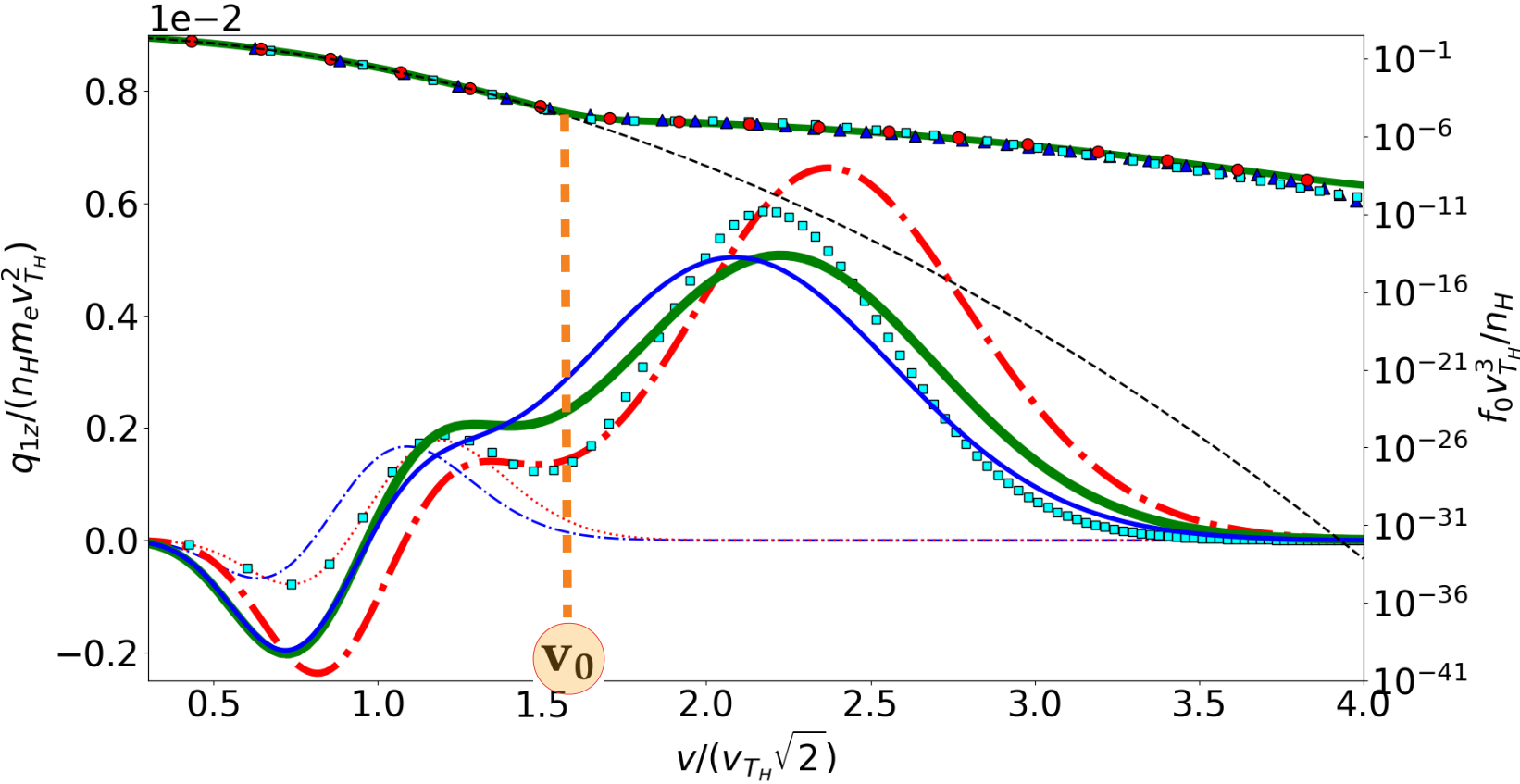}
     \caption{Same as in Fig. \ref{q1NONLOCAL1} for the point at the foot of the temperature gradient marked by a triangle in Fig. \ref{qNonlocal} ($z/\lambda_{ei}^{T_H} \simeq 18.3$).}
    \label{q1NONLOCAL3}
\end{figure*}

Integrands of the heat flux density $q_{1z}= (2\pi m_e/3)v^5f_{1z}$ are shown in Figs. \ref{q1NONLOCAL1}, \ref{q1NONLOCAL2} and \ref{q1NONLOCAL3}. These three figures correspond to the three positions at the temperature profile marked in Fig. \ref{qNonlocal}. Although the overall shapes are similar in a given position, there are shifts induced by the different treatment of electron-electron collisions. 

In the local regime the shape of $q_{1z}$ \eqref{f1SH} remains qualitatively the same everywhere in the temperature profile. Its position on the velocity axis, together with its amplitude, are defined by the local values of temperature and density gradients because the local Maxwellian distribution function \eqref{fM} is entirely specified by these two hydrodynamic quantities. This is not the case in the nonlocal regime. As shown in Figs. \ref{q1NONLOCAL2} and \ref{q1NONLOCAL3}, the shape of $q_{1z}$ is qualitatively different within and beyond the region of maximum temperature gradient. 

The transition between the two shapes occurs in a small region near the foot of the temperature profile. Here, the temperature gradient is smoother and so the diffusive part of the heat flux density is strongly decreased. Because the temperature itself decreases, this diffusive contribution shifts to low velocities. On the contrary, the convective part of the flux carried by fast electrons \cite{gurevich1979thermal} is enhanced, and the position of its maximum is shifted toward higher velocities. This phenomenon has a kinematic origin: to access the cold region an electron must have the appropriate velocity. The further from the region of maximum temperature gradient a region is, the faster the electrons it contains. 

The point in velocity space $v_0$ separating these two contributions corresponds to an abrupt variation of the slope of $f_0$. Beyond the region of maximum temperature gradient, the isotropic part of the distribution function is close to the local Maxwellian distribution function for velocities $v$ smaller than $v_0$, whereas for higher velocities $v \geq v_0$ it is a convolution of the isotropic functions of hotter regions. For $v \leq v_0$ the heat flux density integrand $q_{1z}$ provided by the local theory is not recovered. This is because the local thermodynamic conditions are modified by the presence of delocalized electrons. More precisely, in expression \eqref{f1A} for $\mathbf{f}_1$ the integration is performed from $v$ to $\infty$. So, for a given velocity $v$, the value of $\mathbf{f}_1$ is only determined by $f_0$ for velocities greater than $v$. The convective component ($v \geq v_0$) of $q_{1z}$ is thus independent from the local Maxwellian part of the electron distribution, whereas the component of $q_{1z}$ related to the diffusive part ($v \leq v_0$) is affected by the deformation of $f_0$ for $v \geq v_0$. 

On the contrary, the heat flux density integrand $q_{1z}$ provided by the SNB model is equal to $q_{1, \text{L}}$ for low velocities. This is because the source term in Eq. \eqref{eqDeltaf0} is negligible for $v \ll v_T$. Indeed, the characteristic time for diffusion of the perturbation $\delta f_0$ on a macroscopic length $L$ is $3 L^2 \nu_{ei}^E/v^2 \propto 1/v^5$, whereas its dissipation occurs on a time $1/(\eta \nu_{ee}) \propto v^3$. For $v=v_T$, the latter value is of the same order of magnitude than the time required to reach the quasi-stationary kinetic state described by Eq. \eqref{eqDeltaf0}. For high velocities, the diffusion occurs in a duration much smaller than dissipation, thus leading to significant values of the perturbations. Such a high velocity diffusion corresponds to macroscopic displacements toward cold region within the plasma. They are inhibited by the electric field, which is implicitly accounted for in the diffusion coefficient. For low velocities, the dissipation dominates and the perturbation is only determined by the source term. Therefore, $\delta f_0 \ll f_M$ and $||\delta \mathbf{f}_1|| \ll ||\mathbf{f}_{1, \text{L}}||$ in the low velocity domain.

\subsection{Comparison of code performances}\label{sec43}

To be coupled to a hydrodynamic code, JASMINE must be sufficiently fast. Such an efficiency depends on the degree of simplification that led to the underlying model, but also to its numerical implementation. The latter is detailed in our companion paper \cite{Chrisment2022}. The price for improving the accuracy of the distribution functions is a greater computation time compared to SNB. Without parallelization and for 2000 cells in space, the computation times are given in Tab. \ref{table:1}.

\begin{table}[h!]
\centering
\begin{tabular}{|l|c|r|}
   \hline
   \hline
   Number of groups & \multicolumn{2}{c|}{Computation time [s]} \\
                    & \multicolumn{2}{c|}{SNB \hspace{0.2cm} JASMINE} \\
   \hline
   300 & \multicolumn{2}{c|}{4 \hspace{1.2cm} 13} \\
   1000 & \multicolumn{2}{c|}{13 \hspace{1.1cm} 40} \\
   \hline
   \hline
\end{tabular}
\caption{Computation times required by SNB and JASMINE, without parallelization, for 300 and 1000 velocity groups in the nonlocal case of Sec. \ref{sec4}. The number of cells in space is 2000.}
\label{table:1}
\end{table}

Results shown in the previous section correspond to 1000 groups in velocity. With JASMINE, this number can be reduced to 300 by having a tolerance of about $5 \%$ on the maximum error committed on the heat flux density. With SNB, the number of velocity groups required to obtain a converged heat flux density is lower than for reaching convergence of the distribution function: a few dozen are sufficient and the computation time is a fraction of a second.  

JASMINE - and SNB - provides a stationary kinetic solution for fixed hydrodynamic profiles, whereas ALADIN and OSHUN solve non-stationary kinetic equations. Thus, a rigorous comparison between the computational performances of these codes deserves a special attention, and we only give here an estimate on the order of magnitude with our current implementation: in a macroscopic scale simulation, JASMINE would reduce, at least, the computational time by a factor three with respect to ALADIN, and by a factor thirty with respect to OSHUN.

\section{Conclusion}\label{sec5}

In this paper, we analyzed and improved the model which consists of using the electron-electron operator introduced by Albritton \emph{et al.} \cite{albritton1986nonlocal} together with the P1 closure. The necessity of renormalizing the electron-electron collision frequency has been demonstrated together with an analytical expression of the correction factor. In conditions relevant to ICF, the heat flux calculated with this model agrees with a maximum difference of 10\% with the full kinetic computation, while requiring a much smaller computation time. It is important to correctly describe the first moment of the electron-electron collision operator. Its differential nature is a key element for both computing the electric field properly and capturing the shift toward low velocities of the anisotropic part of distribution function.   

\section*{Acknowledgments}

This work has been done under the auspices of the French Alternative Energies and Atomic Energy Commission (CEA), which the authors thank for its support in the realization of this project.

\section*{Data availability}
The data that support the findings of this study are available from the corresponding author upon reasonable request.

\nocite{*}

\bibliography{apssamp}

\end{document}